\documentclass[twocolumn,
aps,
amsmath,amssymb,
groupedaddress,
superscriptaddress,
longbibliography,
]{revtex4-1}

\usepackage{graphicx}
\usepackage{dcolumn}
\usepackage{bm}
\usepackage[utf8]{inputenc}
\usepackage[T1]{fontenc}
\usepackage[colorlinks=true,citecolor=blue,linkcolor=blue]{hyperref}
\usepackage[nonumberlist, toc, acronym]{glossaries}
	\glsdisablehyper
	\newacronym{fem}{FEM}{finite element method} 
	\newacronym{mst}{MST}{multiple scattering theory}
	\newacronym{kl}{K-L}{Kirchhoff–Love}
	\newacronym{ibz}{IBZ}{irreducible Brillouin zone}
	\newacronym{pc}{PnC}{phononic crystal}
	\newacronym{is}{IS}{inversion symmetry}
	\newacronym{ts}{TRS}{time-reversal symmetry}
	\newacronym{tr}{TR}{time reversal}
	\newacronym{hp}{HP}{homogeneous plate}
	\newacronym{qhe}{QHE}{quantum Hall effect}
	\newacronym{qvhe}{QVHE}{quantum valley Hall effect}
	\newacronym{qshe}{QSHE}{quantum spin Hall effect}
	\newacronym{ti}{TI}{topological insulator}
	\newacronym{dof}{DOF}{degree of freedom}

\begin{document}

\title{Abnormal topological refraction into free medium at sub-wavelength scale\\ in valley phononic crystal plates}

\author{Linyun Yang}
\affiliation{Department of Astronautic Science and Mechanics, Harbin Institute of Technology, Harbin, Heilongjiang 150001, China}
\affiliation{Sorbonne Universit\'e, UPMC Universit\'e Paris 06 (INSP-UMR CNRS 7588),\\ 4, place Jussieu 75005 Paris, France}

\author{Kaiping Yu}
\thanks{Corresponding authors: yukp@hit.edu.cn (Kaiping Yu), bernard.bonello@insp.jussieu.fr (Bernard Bonello)}
\affiliation{Department of Astronautic Science and Mechanics, Harbin Institute of Technology, Harbin, Heilongjiang 150001, China}

\author{Bernard Bonello}
\thanks{Corresponding authors: yukp@hit.edu.cn (Kaiping Yu), bernard.bonello@insp.jussieu.fr (Bernard Bonello)}
\affiliation{Sorbonne Universit\'e, UPMC Universit\'e Paris 06 (INSP-UMR CNRS 7588),\\ 4, place Jussieu 75005 Paris, France}

\author{Wei Wang}
\affiliation{Department of Physics, Hong Kong Baptist University, Kowloon Tong, Hong Kong, China}

\author{Ying Wu}
\affiliation{School of Physics and Optoelectronic Technology, South China University of Technology, Guangzhou, Guangdong 510640, China}

\date{\today}

\begin{abstract}
In this work we propose a topological valley phononic crystal plate and we extensively investigate the refraction of valley modes into the surrounding homogeneous medium. This phononic crystal includes two sublattices of resonators (A and B) modeled by mass-spring systems. We show that two edge states confined at the AB/BA and BA/AB type domain walls exhibit different symmetries in physical space and energy peaks in the Fourier space. As a result, distinct refraction behaviors, especially through an armchair cut edge, are observed. On the other hand, the decay depth of these localized topological modes, which is found to be solely determined by the relative resonant strength between the scatterers, significantly affects the refraction patterns. More interestingly, the outgoing traveling wave through a zigzag interface becomes evanescent when operating at deep sub-wavelength scale. This is realized by tuning the average resonant strength. We show that the evanescent modes only exist along a particular type of outlet edge, and that they can couple with both topological interface states. We also present two designs of topological functional devices, including an elastic one-way transmission waveguide and a near-ideal monopole/dipole emitter, both based on our phononic structure.

\end{abstract}

\maketitle

\newpage

\section{Introduction}

\begin{figure*}[t]
	\centering
	\includegraphics[width=0.75\linewidth]{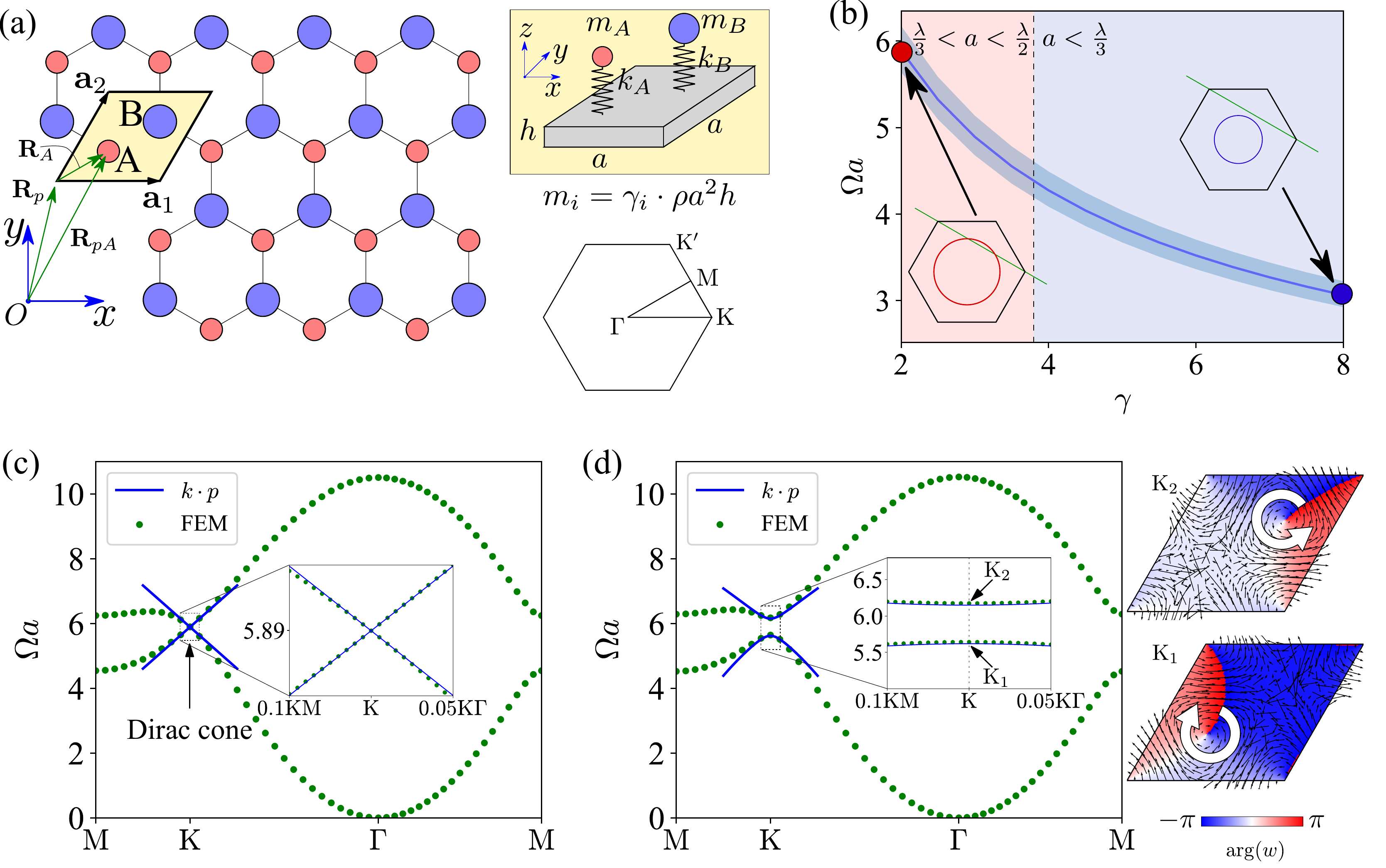}
	\caption{(a) Elastic PnC plate with honeycomb lattice of mass-spring resonators, primitive unit cell and first irreducible Brillouin zone (IBZ). (b) Dirac frequency (solid line) and bulk band gap range (shaded area) at a perturbation level of $\pm 0.1\gamma$ as a function of the average mass $\gamma=(\gamma_A+\gamma_B)/2$. Band structures of the (c) unperturbed $\Delta\gamma=0$ and (d) perturbed $\Delta\gamma=0.2$ PnCs when $\gamma=2$. The right panel of (d) displays the phase distributions and the time-averaged energy flux (black arrows) of the lower and upper gaped states at valley $K$.}
	\label{fig1}
\end{figure*}

Recent advances in topological condensed matter physics have drawn growing attention in the scientific community, due to the excellent performance in wave manipulation, such as the suppression and immunity against sharp bends and defects along the interface or boundary, which is the key feature of topological protection.
The quantum physics originated concepts have been recently extended to the field of classical wave systems through mimicking the \gls{qhe} \cite{haldane2008,lu2014,khanikaev2015,yangzhaoju2015,dingyujiang2019,nash2015,wangpai2015}, the \gls{qshe} \cite{mousavi2015,hecheng2016,meijun2016,SimonYves2017,miniaci2017,Xiabaizhan2017,chaunsali2018,yusiyuan2018,yanglinyun2018} or the \gls{qvhe} \cite{lujiuyang2017,gaofei2018,wuying2018,makwana2018,zhuzhenxiao2019,xieboyang2019,wangwei2019,wangwei2019B,zhangquan2020}.

The realization of classical analogues of \gls{qhe} requires breaking the \gls{ts}, which can be achieved by applying a magnetic field in photonic systems \cite{haldane2008,lu2014}. For acoustic or mechanical systems, active components such as circulating fluid flows \cite{khanikaev2015,yangzhaoju2015,dingyujiang2019} or gyroscopes \cite{nash2015,wangpai2015} have been proposed to create such an effective "magnetic" field to break the \gls{ts}. But this strategy still remains challenging due to the structural complexity. Mimicking \gls{qshe} or \gls{qvhe} in classical waves by breaking the symmetry does not require any external fields. Unlike in electronic crystals, though half integer spin induced Kramers degeneracy is not guaranteed in photonic or phononic systems as they are bosonic, classical analogues of \gls{qshe} can be achieved by carefully designing the system to have two pseudospins, either by integrating the different wave polarization \gls{dof} \cite{mousavi2015,miniaci2017,wangwei2019} or by folding the band structure \cite{hecheng2016,meijun2016,SimonYves2017,Xiabaizhan2017,chaunsali2018,yusiyuan2018,yanglinyun2018} to form a double Dirac cone. In addition, the valley index in solid-state materials can be viewed as a new carrier of information by emulating \gls{qvhe}, originated from the valley \gls{dof} of electrons in graphene \cite{dixiao2007}, can be achieved by breaking the spatial \gls{is} \cite{lujiuyang2017,gaofei2018,wuying2018,makwana2018,zhuzhenxiao2019,xieboyang2019,wangwei2019,wangwei2019B,zhangquan2020}.

Though a variety of topological phases have been demonstrated, most of the current studies concentrate on the robust wave propagation along the topological domain walls inside the lattice structures \cite{khanikaev2015,yangzhaoju2015,dingyujiang2019,nash2015,wangpai2015,mousavi2015,hecheng2016,meijun2016,SimonYves2017,miniaci2017,Xiabaizhan2017,chaunsali2018,yusiyuan2018,yanglinyun2018,lujiuyang2017,wuying2018,makwana2018,wangwei2019B,zhangquan2020,wangwei2020prb}. However, for applicable topological devices in reality, the finite size and boundary effects of the lattices also need to be taken into consideration. Perhaps the most concerned issue is how the topological edge waves would be transmitted into the surrounding homogeneous medium. Various outcoupling effects, such as topological positive/negative, near-zero refraction and wave splitting have recently been reported in valley \gls{pc}s \cite{gaofei2018,zhuzhenxiao2019,xieboyang2019,wangwei2019}, Weyl
\gls{pc} \cite{hehailong2018}, and \gls{qshe} inspired \gls{pc}s \cite{zhangzhiwang2017,huanghongbo2020}.

In this article, we extensively investigate the topological refraction in valley \gls{pc}s. We examine the refraction behaviors of the symmetric and antisymmetric valley modes confined along different domain walls through zigzag and armchair outlet edges, with the influence of both decay depth and Dirac frequency range taken into consideration. Abnormal refraction patterns arise in some cases. When the edge modes are strongly localized, the refracted field spreads in a large area in the background (see Fig.\ref{fig4}(c)), different from what can be estimated from phase matching conditions (see Fig.\ref{fig4}(a)). Likewise, in the sub-wavelength regime, refraction of the expected evanescent waves is forbidden, resulting in strong local resonances and in turn to radiation of elastic energy in the background by the nearly perfect monopoles and/or dipoles formed by the resonators on the edge.

The article is organized as follows: in section
\ref{sec:II}, we present the elastic plate model with honeycomb lattice of mass-spring resonators attached on the top face, and we illustrate the dispersion curves for flexural wave. The $\mathbf{k}\cdot\mathbf{p}$ perturbation method is used to develop the effective Hamiltonian characterizing the band structure and topological properties. We also describe the \gls{mst} for wave field responses under any incident wave. In section \ref{sec:III}, we show the existence of a pairs of edge states confined at the topological domain walls. Various topological refraction patterns are detailed and discussed. In section \ref{sec:IV} we present two examples of possible functional topological devices, including a one-way transmission elastic waveguide and a near perfect monopole/dipole emitter. Finally, the main results of this work and conclusions are summarized in section \ref{sec:V}.

\section{formulation and methods \label{sec:II}}

We consider the flexural wave propagation problem in a \gls{pc} with a honeycomb lattice of mass-spring resonators attached to the top face of an elastic plate, as shown in Fig.\ref{fig1} (a). The lattice constant and the thickness of the \gls{pc} are $a=50~\mathrm{mm}$ and $h=1~\mathrm{mm}$, respectively. The plate material is supposed to be aluminum, with Young's modulus $E=70~\mathrm{GPa}$, Poisson's ratio $\nu=0.3$ and mass density $\rho=2700~\mathrm{kg/m^3}$. Each unit cell consists of two distinct resonators, or "artificial atoms", characterized by their masses $m_{i}$ and stiffnesses $k_{i}$ ($i=A,B$). For simplicity we let all the spring stiffnesses being identical, so the mass term becomes the only varying parameter that enables our system to exhibit different symmetries. Here we define a non-dimensionalized factor to describe the system's mass degree of freedom as
\begin{equation}
	\gamma_i = \frac{m_i}{\rho a^2 h},~i=A,B,
	\label{eq:gamma}
\end{equation}
or equivalently $\gamma_{A,B}=\gamma\mp\Delta\gamma$, where $\gamma=(\gamma_A+\gamma_B)/2$ is the averaged mass and $\Delta\gamma=(\gamma_B-\gamma_A)/2$ is the mass difference which plays the role of a perturbation. The unperturbed system sustains a pair of linearly crossing gapless dispersion curves at each \gls{ibz} corner, forming a Dirac cone which is guaranteed by the $C_{6v}$ and $C_{3v}$ point group symmetries of the \gls{pc} and $K$ point in the reciprocal space according to group theory \citep{lujiuyang2014}. The gapping of the Dirac cone is achieved via a symmetry reduction from $\{C_{6v},~C_{3v}\}$ to $\{C_{6},~C_{3}\}$ \citep{makwana2018} by introducing a small perturbation $\Delta\gamma$ which breaks the spatial \gls{is}.

\subsection{Band structure}

The flexural wave modes in elastic thin plates are characterized by the out-of-plane displacement field $w(\mathbf{r})$. According to \gls{kl} plate theory, the time-harmonic field is governed by
\begin{equation}
	(D\nabla^4 - \rho h\omega^2)w(\mathbf{r}) = f(\mathbf{r}),
	\label{eq:gneq}
\end{equation}
where $D=Eh^3/\big[12(1-\nu^2)\big]$ is the plate flexural rigidity, $\nabla^4=\nabla^2\nabla^2$ is the biharmonic operator, $\omega$ is the angular frequency, and $f(\mathbf{r})$ is the reaction force from the mass-spring resonators,
\begin{equation}
	f(\mathbf{r}) = \sum_{p} \sum_{i=A,B} m_{i}\frac{\omega_{i}^2\omega^2}{\omega_{i}^2-\omega^2}w(\mathbf{R}_{pi})\delta(\mathbf{r}-\mathbf{R}_{pi}),
	\label{eq:f1}
\end{equation}
in which the summation index $p$ labels each elementary cell. This periodic repeating creates the infinite \gls{pc}. $w(\mathbf{R}_{pi})$ represents the flexural wave field at position $\mathbf{R}_{pi}$, $\omega_{i}=\sqrt{k_i/m_i}$ is the resonant frequency of resonator $i$, and $\delta(\mathbf{r})$ is Dirac delta function. In this article we assume that the resonance frequencies of the mass-spring resonators are much higher than the working frequency we are interested in. In the calculation we have set the spring stiffness to be infinite large ($k_i\rightarrow\infty$), which corresponds to the situation that the masses are attached directly to the plate surface. This does not affect our understanding of the mechanisms of the topological transport and refraction into the surrounding \gls{hp}, and keeps our model simple without introducing unnecessary additional resonance features. The corresponding reaction force in Eq.(\ref{eq:f1}) can be further simplified as 
\begin{equation}
	f(\mathbf{r}) = \sum_{p} \sum_{i=A,B} m_{i}\omega^2 w(\mathbf{R}_{pi})\delta(\mathbf{r}-\mathbf{R}_{pi}).
	\label{eq:f2}
\end{equation}

According to the Bloch theory, the eigenmodes of the flexural wave in periodic \gls{pc} plates can be expressed as
\begin{equation}
	w_{n\mathbf{k}}(\mathbf{r}) = \hat{w}_{n\mathbf{k}}(\mathbf{r})\exp(i\mathbf{k}\cdot\mathbf{r}),
	\label{eq:bloch}
\end{equation}
where $\hat{w}_{n\mathbf{k}}$ is the corresponding cell-periodic counterpart of the Bloch state, $\mathbf{k}$ is an arbitrary wave vector in the \gls{ibz} and $n$ labels the eigenfrequency index. To obtain the band structure, Eq.(\ref{eq:gneq}) must be solved in the infinite two dimensional system. Due to the periodicity, Bloch theory Eq.(\ref{eq:bloch}) enables us to limit the solution domain to only one unit cell, together with a quasi-periodic boundary condition $w(\mathbf{r}+\mathbf{a}_j) = w(\mathbf{r})\exp(i\mathbf{k}\cdot\mathbf{a}_j)$. This eigenvalue problem is then solved by using the plate interface of COMSOL Multiphysics, a commercial \gls{fem} software. The mass-spring is modeled by applying frequency-dependent forces using Eq.(\ref{eq:f2}) along vertical axis and an antisymmetric condition is applied on the mid-plane to identify the flexural modes only. Some other approaches for band structure calculations, such as plane wave expansion method, can be found in Refs. \citep{torrent2013,pal2017,chaunsali2018}. It is also useful to introduce a normalized frequency $\Omega a$, where $\Omega$ is defined as
\begin{equation}
	\Omega = \omega\sqrt{\frac{\rho a^2 h}{D}}.
	\label{eq:norm}
\end{equation}

We first calculate the band structure of an unperturbed system with $\gamma=2$, as depicted in Fig.\ref{fig1}(c). As expected, the first two bands linearly cross each other at the normalized frequency $\Omega_D a=5.89$, forming a Dirac cone at the \gls{ibz} corner $K$. The inset shows a comparison between the first principle \gls{fem} calculated dispersion curves (green dots) and the analytical results (blue solid lines) derived from an effective Hamiltonian by $\mathbf{k}\cdot\mathbf{p}$ perturbation method (see section \ref{kp}). Then a $10\%$ level of perturbation $\Delta\gamma=0.2$ is introduced to gap the Dirac cone and open a topological band gap, as plotted in Fig.\ref{fig1}(d). The gapped band structure acts as a local quadratic function of wave vector $\mathbf{k}$ according to our later $\mathbf{k}\cdot\mathbf{p}$ analysis, which accurately captures the band diagrams near $K$ point. The phase distributions and time-averaged energy flux of these two lifted degenerate states ($K_1$ and $K_2$) are displayed in the right panel of Fig.\ref{fig1}(d), which reveals that they can be regarded as valley pseudospin states based on the vortex chirality. The features of valley $K^{\prime}$, the time-reversal counterpart of $K$, can be deduced in a similar way. If we change the sign of mass perturbation, the chiralities of these two lifted states would reverse, which is typically a clue for topological transition. To unambiguously reveal the topological properties of these bands, we use in section \ref{kp} the $\mathbf{k}\cdot\mathbf{p}$ perturbation method to give an analytical expression of the reduced effective Hamiltonian, and obtain the expression of the valley Chern numbers.

It has been demonstrated that the perturbation $\Delta\gamma$ is responsible for the splitting of Dirac cone and the opening of topological band gaps. One may wonder how the average mass $\gamma$ affects the \gls{pc}'s behaviors. In fact, $\gamma$ can be used to tune the Dirac frequency $\Omega_D a$ within a certain range. Actually, it can be seen from Fig.\ref{fig1}(b) that $\Omega_D a$ decreases as $\gamma$ increases. At a certain value of $\gamma$ (vertical dashed line), the line connecting two $K$ points in the reciprocal space is tangent to the equifrequency contour computed in the \gls{hp}. Here this line represents the normal of a zigzag outlet edge of a finite sized lattice. Below this critical $\gamma$ (light red area), the interface normal has intersection points with the equifrequency contour. According to the conservation law of wave vector along the tangent direction of the interface, the direction joining this intersection point to the origin represents the propagation direction of the refracted wave. However, above this critical $\gamma$ (light blue area), we arrive at a deeper sub-wavelength scale  ($a<\lambda/3$) in which the normal to the interface has no intersections with the equifrequency contour. It means that we cannot find a wave vector in the \gls{hp} that fulfills the conservation law of wave vector along the tangent direction of the interface. In other words, the $K$ (or $K^{\prime}$) locked states which propagate inside a lattice cannot refract into a traveling wave into the \gls{hp} when leaving the outlet edge at the current orientation. In section \ref{sec:III} we discuss in detail these different refraction patterns ranging from traveling wave to evanescent wave as the Dirac frequency decreases. 

\subsection{Effective Hamiltonian by $\mathbf{k}\cdot\mathbf{p}$ perturbation \label{kp}}

In this section we show how the biharmonic wave equation Eq.(\ref{eq:gneq}) in the periodic \gls{pc} can be mapped into a massless and massive Dirac Hamiltonian near point $K$ (and $K^{\prime}$ as well) for the unperturbed and perturbed structure, respectively. The explicit form of the reduced effective Hamiltonian will enable us to obtain the expression of valley Chern numbers and unambiguously explain that a nonzero $\Delta\gamma$ contributes to the topological band gap opening.


\textbf{Unperturbed system.} Let us consider the bands near $K$ as an example, i.e., $\mathbf{k}_0=\mathbf{K}$, and that of the $\mathbf{K}^{\prime}$ point can be similarly obtained. Firstly, all the eigenfrequencies $\omega_{n\mathbf{k}_0}$, the corresponding Bloch functions $w_{n\mathbf{k}_0}(\mathbf{r})$ and also their cell-periodic counterpart $\hat{w}_{n\mathbf{k}_0}(\mathbf{r})$ are assumed to be known, which can be numerically obtained from \gls{fem} calculations. Secondly, we expand the cell-periodic counterpart of Bloch mode at $\mathbf{k}$ in terms of $\hat{w}_{n\mathbf{k}_0}$, namely
\begin{equation}
	\hat{w}_{n\mathbf{k}}(\mathbf{r}) = \sum_m A_{nm}(\mathbf{k}) \hat{w}_{m\mathbf{k}_0}(\mathbf{r}),
\end{equation}
in which the expansion coefficients $A_{nm}$ are independent of $\mathbf{r}$ but dependent on $\mathbf{k}$. The reason why we choose to expand $\hat{w}_{n\mathbf{k}}(\mathbf{r})$ into the superposition of $\hat{w}_{n\mathbf{k}_0}(\mathbf{r})$ is that all these functions satisfy the same periodic boundary condition and thus they belong to the same Hilbert space with a proper definition. Then according to Eq.(\ref{eq:bloch}) the Bloch modes at $\mathbf{k}$ can be expanded as
\begin{equation}
	w_{n\mathbf{k}}(\mathbf{r}) = \exp(i\delta\mathbf{k}\cdot\mathbf{r})\sum_m A_{nm} w_{m\mathbf{k}_0}(\mathbf{r}),
	\label{eq:expand}
\end{equation}
where $\delta\mathbf{k}=\mathbf{k}-\mathbf{k}_0$. Inserting this expression and Eq.(\ref{eq:f2}) back into Eq.(\ref{eq:gneq}) yields
\begin{widetext}
	\begin{equation}
		\sum_m A_{nm}\left\{ D\nabla^4+4D\delta\mathbf{k}\cdot i\nabla\nabla^2 -\omega_{n\mathbf{k}}^2\bigg[\rho h+\sum_{i=A,B}m_{0}\delta(\mathbf{r}-\mathbf{R}_{i})\bigg] +\mathcal{O}(|\delta\mathbf{k}|^2)\right\}w_{m\mathbf{k}_0}(\mathbf{r})  = 0.
		\label{eq:gvn_expan}
	\end{equation}
\end{widetext}

Since $\{\omega_{m\mathbf{k}_0},w_{m\mathbf{k}_0}(\mathbf{r})\}$ satisfy the governing equation, and considering the mass-matrix-scaled orthogonality of the Bloch functions
\begin{equation}
	\int_{\mathrm{UC}} \rho^{\prime} h w^*_{n\mathbf{k}_0}(\mathbf{r}) w_{m\mathbf{k}_0}(\mathbf{r}) \mathrm{d}\mathbf{r} = \delta_{m,n},
	\label{eq:orth}
\end{equation}
where $\rho^{\prime}=\rho+1/h\sum_{i=A,B}m_{0}\delta(\mathbf{r}-\mathbf{R}_{i})$, and the integration is performed over one unit cell. So Eq.(\ref{eq:gvn_expan}) can be simplified as
\begin{equation}
	\sum_m [\delta\mathbf{k}\cdot\mathbf{p}_{nm}-(\omega_{n\mathbf{k}}^2-\omega_{n\mathbf{k}_0}^2)\delta_{m,n}] A_{nm}= 0,
	\label{eq:kp}
\end{equation}
where the second and higher order terms of $|\delta\mathbf{k}|$ have been neglected, and
\begin{equation}
	\mathbf{p}_{nm}=4D\int_{\mathrm{UC}} w^*_{n\mathbf{k}_0}\cdot i\nabla(\nabla^2w_{m\mathbf{k}_0}) \mathrm{d}\mathbf{r}.
\end{equation}

As we are interested in the dispersion around the Dirac cone, and the other bands are far away from the Dirac frequency (Fig.\ref{fig1}), the infinite summation in Eqs.(\ref{eq:gvn_expan}) and (\ref{eq:kp}) can be truncated and only the two degenerate modes are conserved. Therefore $\mathbf{p}$ is reduced into a 2-by-2 matrix in which each entry itself is a two dimensional vector. The reduced Hamiltonian has therefore the following form
\begin{equation}
\begin{aligned}
	\mathcal{H} &= \delta\mathbf{k}\cdot
	\begin{bmatrix}
		\mathbf{p}_{11} & \mathrm{Re}(\mathbf{p}_{12}) +i\mathrm{Im}(\mathbf{p}_{12}) \\
		\mathrm{Re}(\mathbf{p}_{12}) -i\mathrm{Im}(\mathbf{p}_{12}) & -\mathbf{p}_{11}
	\end{bmatrix} \\[2ex]
	&= \delta\mathbf{k}\cdot[\mathrm{Re}(\mathbf{p}_{12})\sigma_x-\mathrm{Im}(\mathbf{p}_{12})\sigma_y+\mathbf{p}_{11}\sigma_z],
\end{aligned}
\label{eq:eff_H0}
\end{equation}
where we have used the Hermitian property of $\mathbf{p}$, in addition to $\mathbf{p}_{22}=-\mathbf{p}_{11}$, obtained from point group symmetry analysis \citep{makwana2018}, and $\sigma_x,~\sigma_y,~\sigma_z$ are the Pauli matrices. Eq.(\ref{eq:eff_H0}) has the same form as that in Ref. \citep{meijun2012} for acoustic systems, while here we are dealing with flexural waves in structured plates. The $x$ and $y$ components of $(\mathrm{Re}\mathbf{p}_{12},-\mathrm{Im}\mathbf{p}_{12},\mathbf{p}_{11})$
construct two three-dimensional vectors in the pseudo-spin space, and they should have the same length as required by the isotropy of the Dirac cone. Note that the choices of the orthogonal degenerate eigenmodes in Eq.(\ref{eq:expand}), related by unitary transformations, are not unique, so the matrix $\mathbf{p}$ is dependent on the specific choice of the basis. In order to map it into a Dirac Hamiltonian, \citet{meijun2012} proposed a scheme by applying proper rotations of the axes in the pseudo-spin space, which cancels out the $\mathbf{p}_{11}$ term. In fact, these rotations correspond to a unitary transformation of the selected basis eigenfunctions, in this article we achieve the vanishing of $\mathbf{p}_{11}$ by an optimization procedure. Our numerical results show that $\mathbf{p}_{12}=(A,-iA)$ for $K$ and $\mathbf{p}_{12}=(-A,-iA)$ for $K^{\prime}$, where $A$ is a real number ($A=4.277\times10^4$ in our calculation), so we have
\begin{equation}
	\mathcal{H} = A(\tau\delta k_x\sigma_x+\delta k_y \sigma_y),
	\label{eq:14}
\end{equation}
where $\tau=\pm 1$ denotes the valley degree of freedom, corresponding to the $K$ and $K^{\prime}$ valleys, respectively. The eigenvalues of $\mathcal{H}$ are
\begin{equation}
	\omega_{n\mathbf{k}}^2-\omega_{D\mathbf{k}_0}^2 = \pm A|\delta k|,
\end{equation}
where $\omega_{D\mathbf{k}_0}$ is the Dirac frequency. The above expression implies that $A$ represents the gradient of eigenvalues around the Dirac cone,
\begin{equation}
	A = \left|\frac{\Delta (\omega^2)}{\delta k}\right| \equiv v_D.
\end{equation}

\textbf{Perturbed system.} When a nonzero mass perturbation $\Delta\gamma$ is added to the \gls{pc}, the Dirac cone would be lifted. To develop an effective Hamiltonian for this perturbed system, we also start by expanding $w_{n\mathbf{k}_0}$ in terms of the two degenerate states as we did in the unperturbed case, which means Eqs.(\ref{eq:expand}) and (\ref{eq:orth}) still remain valid. The effect of $\Delta\gamma$ (or $\Delta m$) takes place in the mass term in Eq.(\ref{eq:gvn_expan}), so a correction $m_0\rightarrow m_i$ needs to be applied. After a similar procedure as we did in the previous section, an extra term emerges in Eq.(\ref{eq:kp}) as
\begin{equation}
	\sum_m [\delta\mathbf{k}\cdot\mathbf{p}_{nm}+ M-(\omega_{n\mathbf{k}}^2-\omega_{n\mathbf{k}_0}^2)\delta_{m,n}] A_{nm}= 0,
\end{equation}
where the extra term reads
\begin{equation}
	M=\Delta m \omega_{n\mathbf{k}_0}^2\cdot[ w_{m\mathbf{k}_0}^*(\mathbf{r})w_{n\mathbf{k}_0}(\mathbf{r})]\big|_{\mathbf{r}=\mathbf{R}_B}^{\mathbf{R}_A}.
\end{equation}

We found that $M$ is a diagonal matrix and the nonzero entries have opposite signs, i.e., $M=\Delta M\sigma_z$. Therefore the effective Hamiltonian for the perturbed system is 
\begin{equation}
	\mathcal{H}_p = v_D(\tau\delta k_x\sigma_x+\delta k_y \sigma_y) +  \Delta M\sigma_z,
	\label{eq:effH}
\end{equation}
which is exactly the form of a massive Dirac Hamiltonian. The eigenvalues of $\mathcal{H}_p$ are given by
\begin{equation}
	\omega_{n\mathbf{k}}^2-\omega_{D\mathbf{k}_0}^2 = \pm \sqrt{v_D^2\delta k^2+\Delta M^2}.
\end{equation}

By letting $\delta\mathbf{k}=0$, it becomes obvious that the effective mass $|\Delta M|$ has a clear physical meaning, that is the half band gap size of the eigenvalues $(\omega^2_{2\mathbf{k}_0}-\omega^2_{1\mathbf{k}_0})/2$. As shown in Fig.\ref{fig1} (c) and (d), good agreements between the band structures predicted by the effective Hamiltonian and the results from
first principle \gls{fem} calculations can be found around the valley points. We stress that the parameter $A$ in Eq.(\ref{eq:14}) and Dirac mass term are both derived from the two degenerate eigenfunctions, and that no data fitting is involved here.
 
From the effective Hamiltonian in Eq.(\ref{eq:effH}), the local Berry curvature of the lowest band at valleys $K$ and 
$K^{\prime}$ can be analytically evaluated as \cite{dixiao2010,wangwei2019B}
\begin{equation}
	B_{K,K^{\prime}} = \pm\frac{\Delta Mv_D^2}{2(v_D^2\delta k^2+\Delta M^2)^{3/2}}.
\end{equation}

Because of the \gls{pc}'s \gls{ts} is not broken, the total Chern number over the entire \gls{ibz} vanishes and the global nontrivial topology is supported. However, the nonzero valley Chern number obtained by integrating the Berry phase in a half \gls{ibz} area consisting a single valley as \cite{dixiao2010}
\begin{equation}
	C_{K,K^{\prime}} = \frac{1}{2\pi}\int B(\mathbf{k})\mathrm{d}^2\mathbf{k}=\pm\frac{1}{2}\mathrm{sgn}(\Delta \gamma),
\end{equation}
which supports a good quantized topological index. According to the bulk-edge correspondence, topological protected nontrival edges state would emerge at the domain wall between two lattices with opposite valley Chern numbers for each valley.

\begin{figure*}[t]
	\centering
	\includegraphics[width=0.9\linewidth]{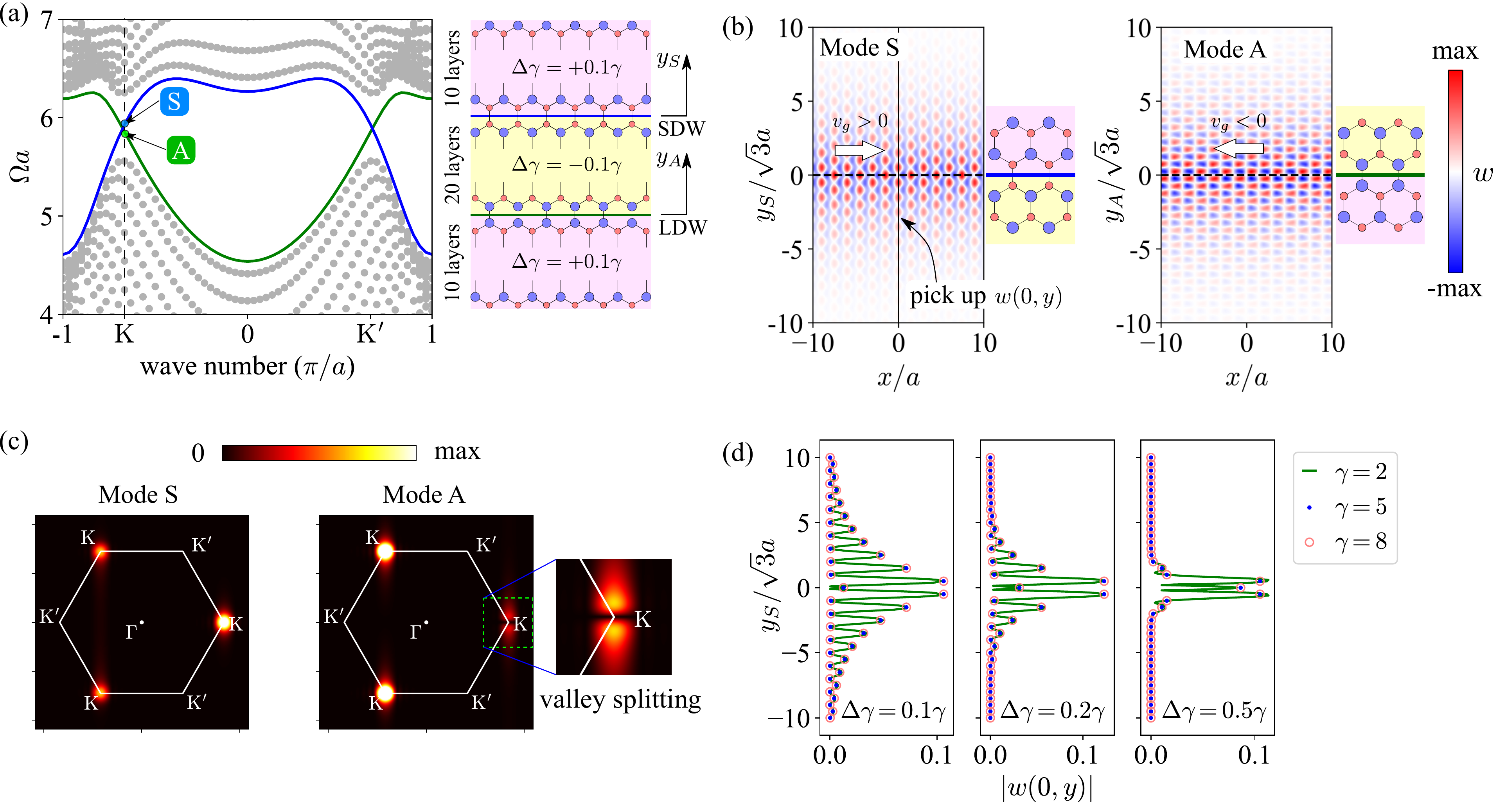}
	\caption{(a) Band structure of a supercell consisting of 20 layers with $\gamma=2$ and $\Delta\gamma=-0.1\gamma$ type of \gls{pc}s sandwiched in between two domains made of 10 layers with $\gamma=2$ and $\Delta\gamma=+0.1\gamma$. (b) The eigenmodes and (c) the corresponding Fourier spectra of the symmetric (S) and antisymmetric (A) valley topological states. (d) Displacement magnitudes of mode S along the perpendicular direction of the topological interface for various average masses $\gamma$ and perturbation levels $\Delta\gamma$.}
	\label{fig2}
\end{figure*}

\subsection{Multiple scattering for finite cluster of resonators}

Before we further investigate the topological transport of valley modes, we introduce the \gls{mst} for the total wave field calculation under any type of incident wave fields, which is frequently used in our later simulations. We can rewrite Eq.(\ref{eq:gneq}) as
\begin{equation}
	\nabla^4 w(\mathbf{r}) - \kappa^4 w(\mathbf{r})
	= \sum_{\alpha}t_{\alpha}w(\mathbf{R}_{\alpha})\delta(\mathbf{r}-\mathbf{R}_{\alpha}),
	\label{eq:gneq2}
\end{equation}
where $\kappa=(\rho h\omega^2/D)^{1/4}=\sqrt{\Omega/a}$ is the wave number in the \gls{hp}, $\alpha$ is a label of the resonators, and $t_{\alpha}$ measures the resonant strength, which is given by

\begin{equation}
	t_{\alpha} = \frac{m_{\alpha}\omega^2}{D}
	=\gamma_{\alpha}\Omega^2
	\label{eq:tpn}
\end{equation}

For truncated finite clusters of resonators, the total wave field distribution under an external incident wave can be semi-analytically solved using \gls{mst} method: each resonator acts like a point source and radiates a cylindrical wave. The total field is thus the superposition of all the scattered fields and external incident field,
\begin{equation}
	w(\mathbf{r}) = w_{\mathrm{inc}}(\mathbf{r}) + \sum_{\alpha}t_{\alpha}w(\mathbf{R}_{\alpha}) G(\mathbf{r};\mathbf{R}_{\alpha}),
	\label{eq:totw}
\end{equation}
where $w_{\mathrm{inc}}(\mathbf{r})$ is the incident field, and $G(\mathbf{r};\mathbf{R}_{\alpha})=i/(8\kappa^2)[H^{(1)}_0(\kappa |\mathbf{r}-\mathbf{R}_{\alpha}|)+2i/\pi K_0(\kappa |\mathbf{r}-\mathbf{R}_{\alpha}|)]$ is the Green's function which satisfies $\nabla^4 w(\mathbf{r}) - \kappa^4 w(\mathbf{r})
=\delta(\mathbf{r}-\mathbf{R}_{\alpha})$ \citep{torrent2013}. Considering the total field at an arbitrary resonator's position $\mathbf{r}=\mathbf{R}_{\beta}$, from Eq.(\ref{eq:totw}) we have the linear system
\begin{equation}
	\sum_{\alpha}\left[\delta_{\alpha,\beta}-t_{\alpha}G(\mathbf{R}_{\beta};\mathbf{R}_{\alpha})\right]w(\mathbf{R}_{\alpha}) = w_{\mathrm{inc}}(\mathbf{R}_{\beta}),
\end{equation} 
from which we can derive the displacement of each resonator. The total field can then be constructed by using Eq.(\ref{eq:totw}). This semi-analytical Green's function method constitutes an ideal approach for efficient calculations of the total wave field response under any given incident wave. For example, with a personal computer, it takes less than one minute to simulate the system in Fig.\ref{fig5} (e) featuring $3416$ resonators.

\section{Topological protected refraction of valley states \label{sec:III}}

\subsection{Topological edge bands and valley-locked states}

\begin{figure*}[t]
	\centering
	\includegraphics[width=0.9\linewidth]{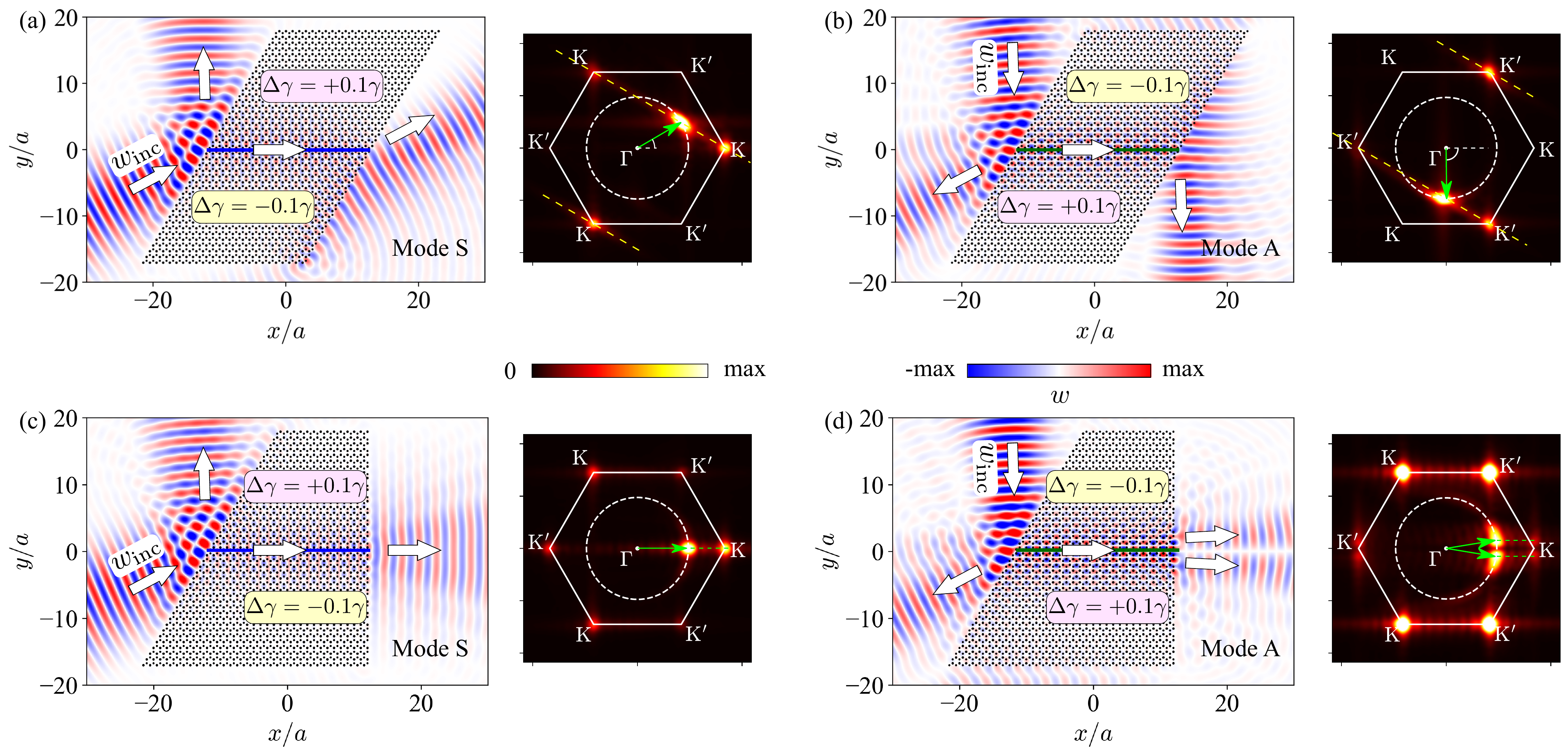}
	\caption{Topologically protected refraction of valley states into the surrounding \gls{hp} through different outlet edges and the corresponding Fourier spectra. The refraction of (a) symmetric and (b) antisymmetric valley states through a zigzag interface. (c), (d) the same as (a) and (b) but with an armchair interface. The Fourier transformations are performed in a square region of $[0,20a]\times[-10a,10a]$, and the structural parameters are $\gamma=2,~\Delta\gamma=\pm 0.1\gamma$.}
	\label{fig3}
\end{figure*}

To investigate the edge states, we consider a supercell made of 20 unit cells with negative perturbation $\Delta\gamma=-0.2$ sandwiched in between 10 units with positive perturbation $\Delta\gamma=+0.2$ on each side, while the average mass is set to $\gamma=2$. A schematic view of the supercell, together with the band structure along $k_x$ direction are shown in Fig.\ref{fig2} (a). Clearly, there are two types of zigzag domain walls, one is constructed with two small neighboring masses (termed as "SDW") and the other one features two large ones ("LDW"). The projected band diagram, which can be obtained by applying periodic boundary conditions on the left/right and top/bottom edge pairs, unambiguously shows that there are two branches of edge states located in the reopened bulk band gap range (blue and green bands in Fig.\ref{fig2}(a)). In Fig.\ref{fig2} (b) we plot the eigenmodes at $k_x=-2\pi/3a$ (valley $K$) of the two branches, both of which the displacements are confined near the topological domain walls, and decay rapidly as the distances from the domain walls increase.

One may ask which factor(s) dominates the displacement decay into the bulk. To answer this question, we systematically calculate the projected band structures and the eigenmodes for various combinations of the structural parameters $(\gamma,\Delta\gamma)$. Then the displacement magnitudes along a perpendicular direction of the domain wall are extracted and displayed in Fig.\ref{fig2}(d), taking the blue branch band as an example. We consider three values of the average mass ($\gamma=2,5,8$), and for each of them we introduce three levels of perturbation ($\Delta\gamma/\gamma=0.1,0.2,0.5$). The nine corresponding curves are arranged in such a way that the results of different average masses but with the same perturbation level are grouped together and are plotted in the same subfigure for better comparison. As we can see, in each panel, the three curves (solid green line, blue dots and red circles) almost coincide with each other. It indicates that for a fixed perturbation level, the influence of the values of $\gamma$ can be nearly neglected. Then by comparing the decay depths between different panels, it can be seen that the larger the perturbation level is, the more localized the topological mode will be. Our analysis reveals that the decay depth is not solely determined by $\gamma$ or $\Delta\gamma$ individually, but by the ratio $\Delta\gamma/\gamma$.

Besides the locally confinement of the displacement distributions, another interesting and important feature of these two edge states, is the symmetry of the mode shape. As displayed in Fig.\ref{fig2}(b), the eigenmode localized at the SDW is symmetric about the domain wall, and therefore is labeled as mode S, while the other one is antisymmetric (mode A) about the domain wall. In order to get a better understanding on these two eigenmodes, spatial Fourier transformations are performed. The result in Fig.\ref{fig2}(c) shows that there are three highlighted peaks at $K$ points $[(4\pi/3a,0),(-2\pi/3a,\pm 2\pi/\sqrt{3}a)]$ in the \gls{ibz} for mode S. However, for mode A, the peak near the horizontal $K$, i.e., $(4\pi/3a,0)$ is not exactly at that point, but is splitted into two weaker out-of-phase peaks with a small shift along $\pm k_y$ direction. The peak splitting can be ascribed to the antisymmetry nature of mode A, which forces the vanishing of the peak at horizontal $K$, because it represents a symmetric plane wave component about the domain wall. This suggests that mode A can operate as a splitter when the wave leaves the \gls{pc} through an armchair outlet edge as will be discussed below.

\subsection{Topological refraction through zigzag and armchair interfaces}

We consider in this section several finite clusters of resonators with different cut edges (zigzag or armchair interfaces) embedded in an infinite \gls{hp}, forming  finite sized topological \gls{pc}s, and we study the refraction of the topological valley states upon transmission into the surrounding medium. We use a Gaussian beam with a proper incident angle emitted from the left edge of the \gls{pc} to excite the topological valley states. The structural configurations are still chosen to be $\gamma=2,~\Delta\gamma=0.2$, and the operating frequency is near the Dirac degenerate point $\Omega a=5.89$. The corresponding equifrequency contour is shown as dashed circles in the Fourier space. In Fig.\ref{fig3}(a)-(d) we plot the \gls{mst} simulated wave fields and their corresponding Fourier spectra. The Fourier transforms are performed in the region $[0,20a]\times[-10a,10a]$, which simultaneously includes the topological valley states and the refracted waves.

The \gls{pc}s in Fig.\ref{fig3}(a) and (c) are constructed by positive $\Delta\gamma=0.2$ type lattice above a lattice with negative $\Delta\gamma=-0.2$, forming a SDW which supports symmetric valley modes. For the zigzag interface case displayed in Fig.\ref{fig3}(a), the right going waves suggests that we have a positive group velocity, so from the band structure we know that $K$ valley-locked states are excited, as confirmed by the three peaks in the Fourier spectrum. The refraction direction can be predicted according to the conservation of tangent components of wave vector parallel to the interface. As can be seen in Fig.\ref{fig3}(a)-(d), the predicted refraction angle perfectly matches with the simulated refracted wave. It should be noted that besides a traveling refracted wave emitted into the surrounding medium, there is also a surface wave propagating along the lower half edge corresponding to the $K(-2\pi/3a,-2a/\sqrt{3}a)$ component. As for the case of armchair interface in Fig.\ref{fig3}(c), we also obtain a good agreement between the equifrequency analysis and the wave field. In this case both $K$ and $K^{\prime}$ states are excited, which reveals that the armchair interface causes strong reflection of valley modes.

By inverting the top and bottom lattices in Fig.\ref{fig3}(a) and (c), we obtain the structures shown in 
Fig.\ref{fig3}(b) and (d), that both feature a topological interface of type LDW and support antisymmetric modes. However, for the antisymmetric branch, the positive group velocity implies we are dealing with the $K^{\prime}$-locked states. In Fig.\ref{fig3}(b) the wave vector conservation law leads to a refraction angle different from the case of symmetric mode. Again, perfect agreement with the simulated wave beam is obtained. As for the additional evanescent mode, it propagates along the upper half of the edge, corresponding to the $(2\pi/3a,2a/\sqrt{3}a)$ $K^{\prime}$ states. Similar to Fig.\ref{fig3}(c), the antisymmetric valley mode also encounters strong inter valleys interactions, induced by the reflection at the armchair interface. The outgoing refraction waves are composed of two out-of-phase beams, resulting from the valley splitting mechanism as illustrated in Fig.\ref{fig2}(c).

\subsection{Influence of decay depth on the refracted waves}

\begin{figure}[t]
	\centering
	\includegraphics[width=\linewidth]{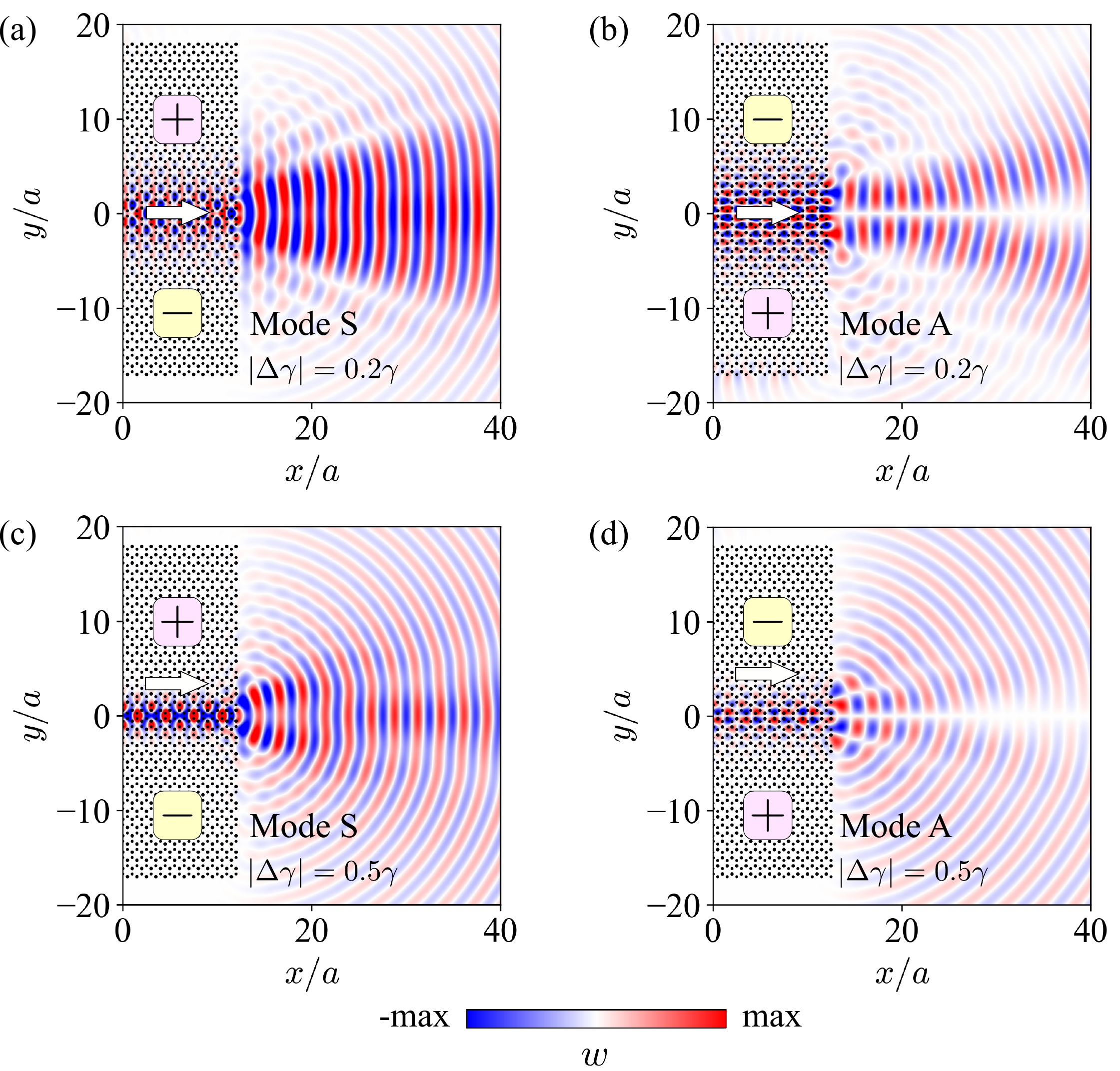}
	\caption{Refracted patterns of topological S and A valley modes with different decay depths ($\gamma=2,~|\Delta\gamma|/\gamma=0.2,0.5$).}
	\label{fig4}
\end{figure}

From the comparison between the different panels in Fig.\ref{fig2}(d), it can be seen that for a given value of $\gamma$, a larger perturbation ratio $\Delta\gamma/\gamma$ results in a shorter decay depth of the topological states. How the variation of the decay depth influences the refracted waves is discussed in this section. Since the Dirac frequency is not altered as $\gamma$ is remained fixed, one may naively think that the only effect is the change of the width of the refracted beam. However, as the decay depth decreases, the topological states becomes increasingly localized within a narrow channel. In the extremely short decay depth case, only a few resonators near the topological domain wall are vibrating while the other ones on the outlet edge have almost no displacements. As a result, strong local resonance occurs at the outlet port and the refracted wave is like a field radiated from a point-like source.

To confirm our interpretation, we examine the wave refracted field resulting from different perturbation levels ($\Delta\gamma/\gamma=0.2,~0.5$) while keeping $\gamma=2$ fixed. The results for both symmetric and antisymmetric valley modes are displayed in Fig.\ref{fig4}. Note that Fig.\ref{fig3}(c) and (d) also illustrate the case of $\gamma=2,\Delta\gamma/\gamma=0.1$ for S and A modes, respectively. Let us have a look on the symmetric valley modes, as the perturbation level increases from $0.1$ to $0.5$. We can see that the excited valley state progressively narrows down, and that the refracted wave gradually changes from a plane wave to a monopole-like radiated field. Similar phenomenon can be found as we vary the perturbation level for the antisymmetric valley state, with the only difference that the refracted wave is a dipole-like source radiated field at large $\Delta\gamma/\gamma$ values. In that case it is the result of the out-of-phase vibrations of A valley state on both sides of the domain wall.

\subsection{Influence of Dirac frequency on the refracted waves}

\begin{figure*}[t]
	\centering
	\includegraphics[width=0.8\linewidth]{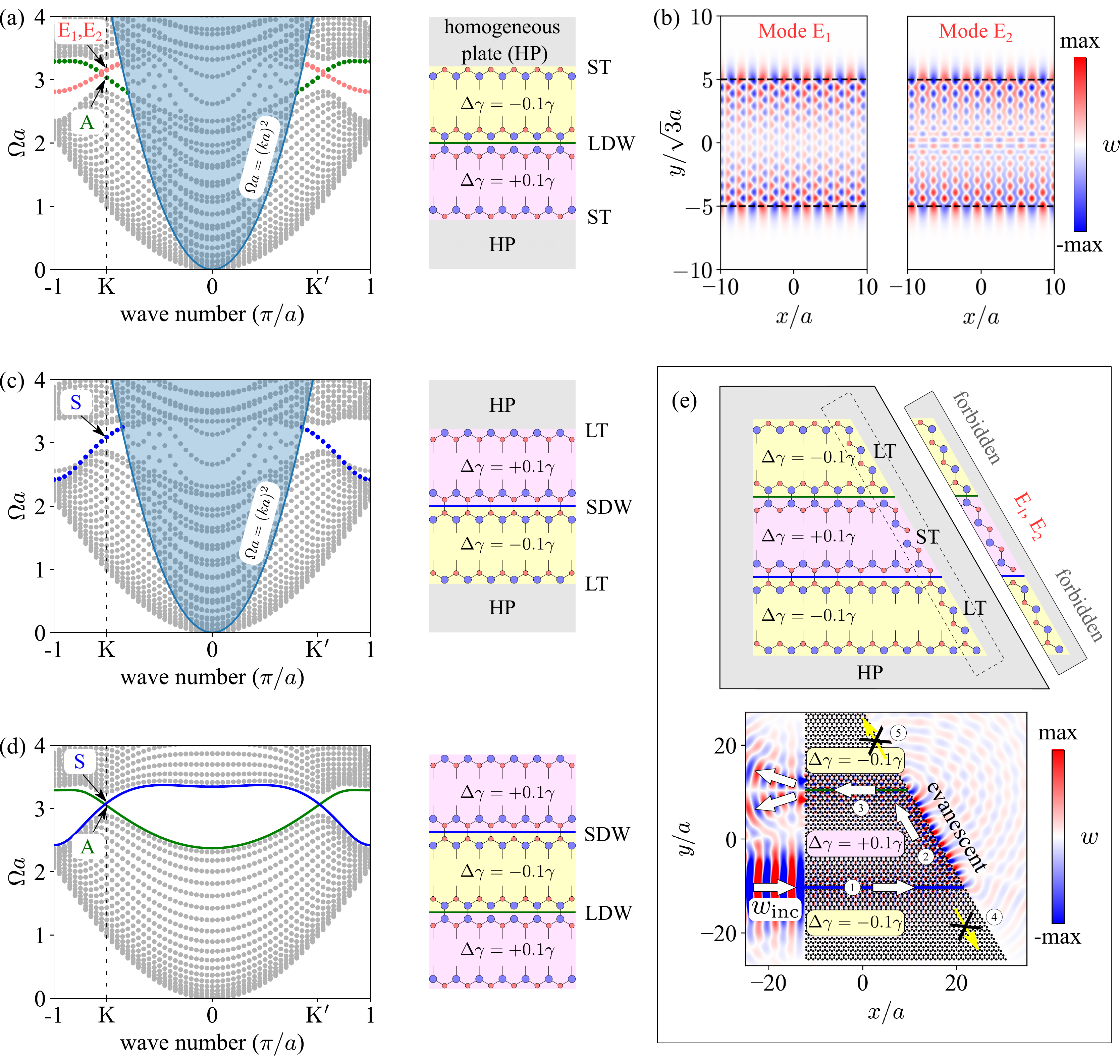}
	\caption{(a) Band structure of a suppercell with ST type edges. (b) The out-of-plane displacement field of eigenmodes E1 and E2. (c) The same as (a) but with LT type edges. (d) Reference band structure of the pure topological \gls{pc}. (e) Coupling between topological valley edge states and evanescent modes in the deep sub-wavelength regime ($\gamma=8,~\Delta\gamma=\pm 0.1\gamma$).}
	\label{fig5}
\end{figure*}

We have shown in section \ref{sec:II} that various Dirac frequencies can be achieved by tuning the average mass $\gamma$ (see 
Fig.\ref{fig1}(b)). Comparing the equifrequency contour in the \gls{hp} and the \gls{ibz} at a particular Dirac frequency, we find that when
\begin{equation}
	\frac{2\pi}{\lambda} > \frac{2\pi}{3a}, \text{~or~} a>\frac{\lambda}{3},
\end{equation}
there is \textit{one} wave vector in the \gls{hp} whose parallel component along a $60^{\circ}$ interface matches with the $K$ (or $K^{\prime}$) locked topological edge states (see the red region in Fig.\ref{fig1}(b)). This wave vector normally gives the direction of the refracted wave, as demonstrated in the previous sections. As $\gamma$ increases, the equifrequency contour shrinks gradually. One can imagine that the refraction angle will finally converge to $90^{\circ}$, and beyond a certain value of $\gamma$, no such a wave vector can be found (see the blue region in Fig.\ref{fig1}(b), $a<\lambda/3$) and therefore only evanescent modes can exists along the interface and no traveling waves will be emitted into the \gls{hp}.

Now we chose a different structural configuration $(\gamma=8,|\Delta\gamma|=0.1\gamma)$, as marked by the blue dot in Fig.\ref{fig1}(b), which allows us to work at a deeper sub-wavelength scale (Dirac frequency $\Omega_D a=3.06$, wavelength $\lambda=3.59a$) and prohibits the refraction of a traveling wave into the \gls{hp}. To have a better understanding of this phenomenon, we calculate the band structures of a suppercell constructed by the \gls{pc} and \gls{hp}, and check the possibility of the existence of surface waves, just as we did for the topological valley states. There are two different types of interface between the \gls{pc} and the \gls{hp}. One is with the small resonators at the very end of the \gls{pc}'s cut edge, shown in Fig.\ref{fig5}(a) and labeled as "ST", and the other one (termed as "LT") is with the large resonators at the outmost edge, as shown in Fig.\ref{fig5}(c). As a reference, the band structure for the pure topological \gls{pc} is also presented in Fig.\ref{fig5}(d). One can find that the topological states are moved down to around $3.06$ because of the choice of a larger $\gamma$.

For the first supercell structure in Fig.\ref{fig5}(a), there are one LDW topological interface and two ST edges. The calculated band structure shows that three edge states emerge in the bulk gap range which includes: one antisymmetric topological valley mode (green dots) localized near the LDW and two degenerate surface modes (namely E1 and E2, marked by red dots) localized near the ST edges. The eigenmodes of these two degenerate states are displayed in Fig.\ref{fig5}(b), which clearly shows that the displacement decays rapidly as the measurement point goes away from the edges. There are two points about the evanescent states (E1, E2) that deserve special attention: (1) they have the same group velocity but different mode shapes (being symmetric and antisymmetric about the LDW, respectively), (2) both of the two evanescent modes and the topological A mode at valley $K$ are all below the acoustic line. The latter property implies that there is no possibility to find a mode in the \gls{hp} that can couple with the valley-locked topological states or the evanescent modes.

The band structure analysis corresponding to the LT edge is displayed in Fig.\ref{fig5}(c). Only one branch (blue dots) in the bulk gap exists, namely the symmetric topological state localized near SDW. Unlike the aforementioned ST edge, no evanescent modes are found in the gap.

Let us consider now a trapezoid shape \gls{pc} constructed by $24$ layers of positive perturbed \gls{pc} sandwiched in between $18$ layers of negative perturbed lattice at both upper and lower sides, as shown in Fig.\ref{fig5}(e). In this \gls{pc} both SDW (path 1) and LDW (path 3) topological interfaces, and both ST (path 3) and LT (path 4 and 5) types of edges are included. A horizontal Gaussian beam impinges the \gls{pc} at the port of path 1, which excites the symmetric valley topological mode. When the wave arrives at the right end of the \gls{pc}, it turns into an evanescent wave and propagates along path 2 with a sharp bend, but not along path 4, although the bend is more gentle. When the wave reaches the crossing point of path 3 and 5, instead of keeping going straight along path 5, the wave turns its direction with a sharp bend, and transforms into the antisymmetric valley topological mode along path 3. Finally the topological state refracts into two out-of-phase beams when leaving the \gls{pc} at the armchair outlet edge. The above \gls{mst} simulation of the wave propagation illustrates the efficient coupling between the topological valley states and the evanescent waves whose existence depends on the type of the edge.


\section{Examples of functional devices designs \label{sec:IV}}

\begin{figure}[t]
	\centering
	\includegraphics[width=\linewidth]{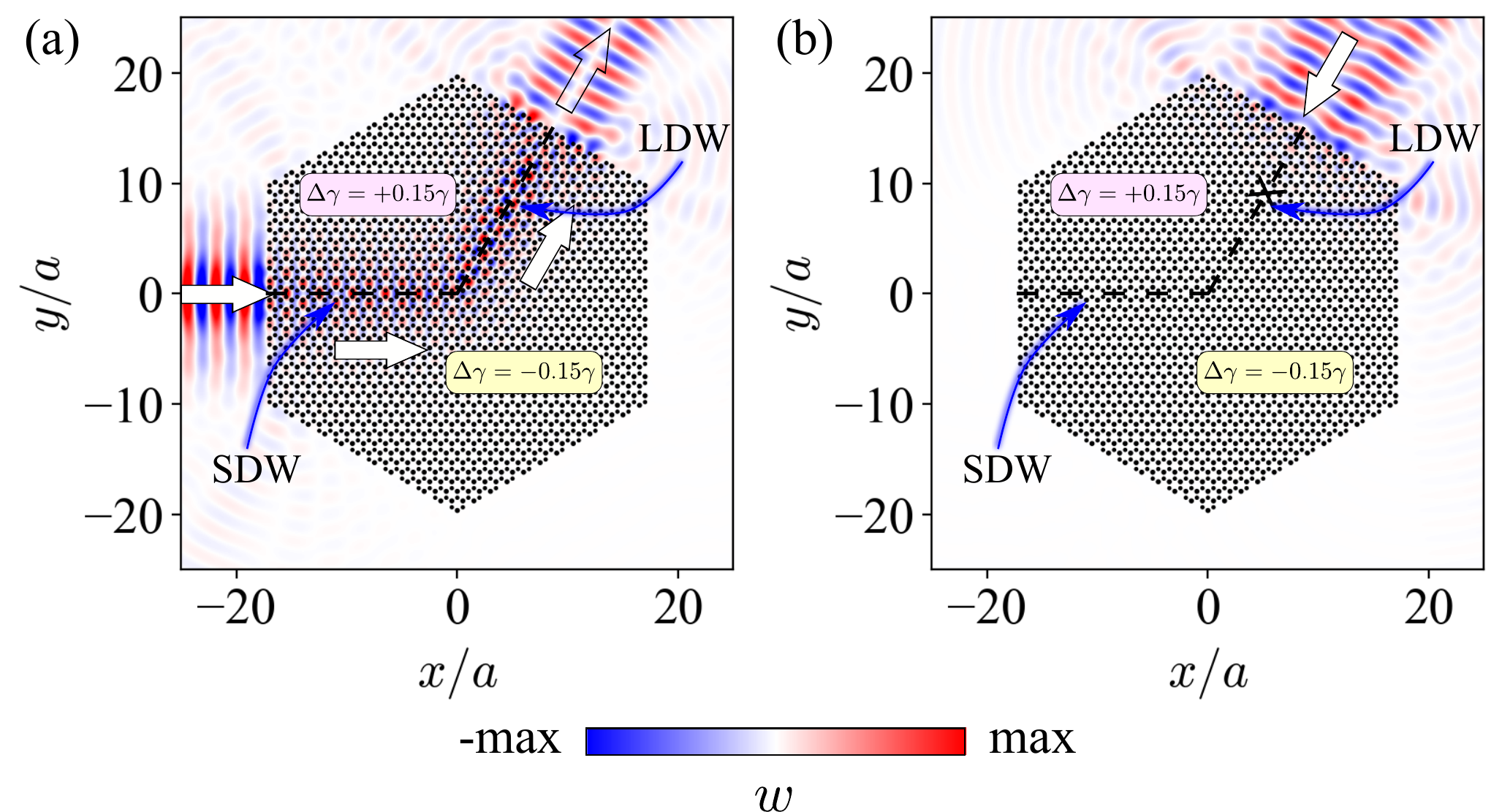}
	\caption{Topological elastic one-way transmission waveguide based on a valley selective excitation ($\gamma=2,|\Delta\gamma|/\gamma=0.15$).}
	\label{fig6}
\end{figure}

\begin{figure}[t]
	\centering
	\includegraphics[width=\linewidth]{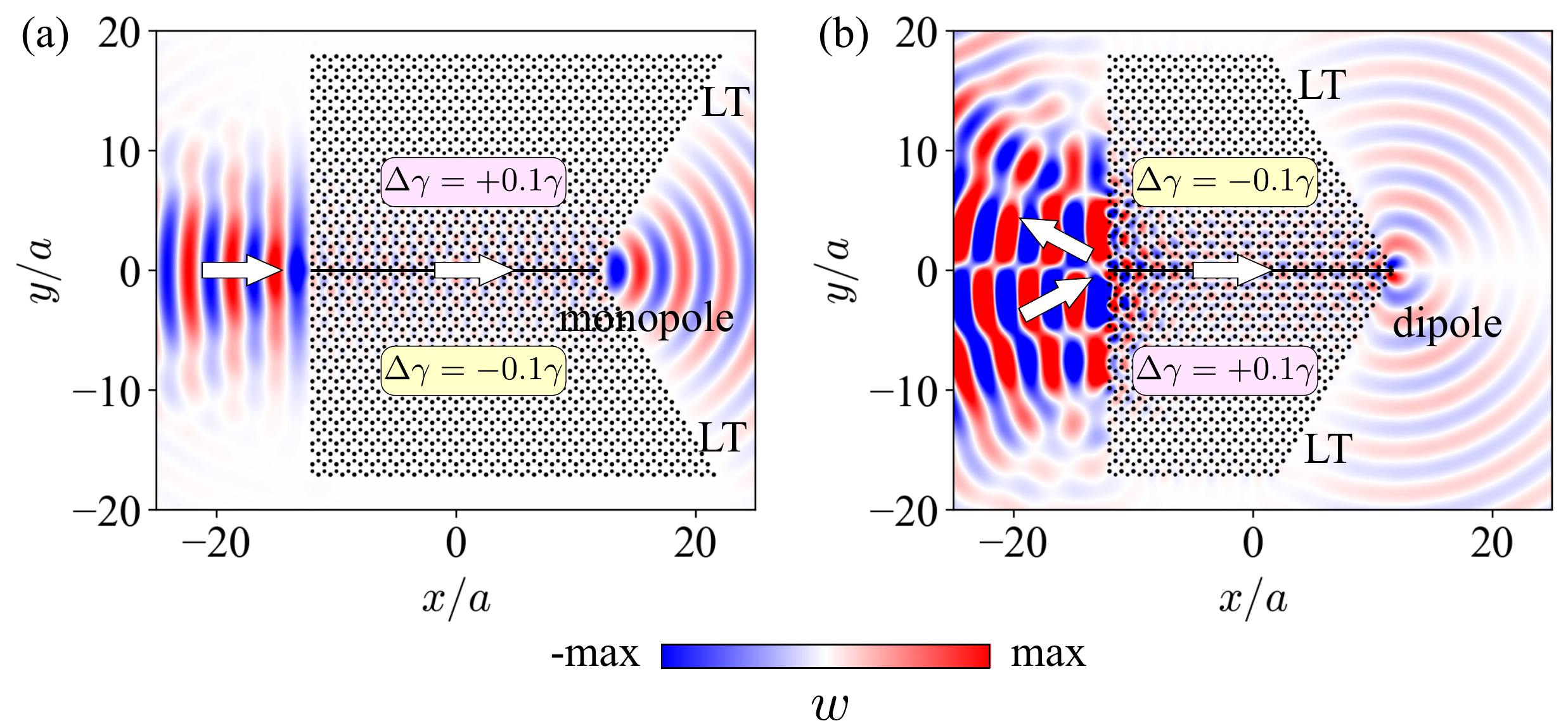}
	\caption{Nearly ideal topological monopole and dipole emitters at the sub-wavelength scale ($\gamma=8,~\Delta\gamma=\pm 0.1\gamma$).}
	\label{fig7}
\end{figure}

We describe in this section two examples of functional designs of topological devices based on our findings. The first one is a topological elastic diode, which allows transmission of elastic wave along only one direction. It is inspired by the different symmetries of the valley modes localized near SDW and LDW, resulting in the selective excitation \citep{lujiuyang2017} of these two modes. The structure of this \gls{pc} is shown in Fig.\ref{fig6}, featuring a topological wave guide formed by two distinct valley phases. The horizontal and then oblique interfaces are of type SDW and LDW and support the symmetric and antisymmetric valley modes, respectively. From the refraction patterns in Fig.\ref{fig3}(c) and (d) and considering the reciprocity, we know that a normally incident Gaussian beam can only excite the symmetric mode, but not the antisymmetric one. Our simulations in Fig.\ref{fig6} clearly show the one-way property of this \gls{pc}.

The second interesting application is to construct a topological monopole or dipole emitter. We chose the same structural parameter as in Fig.\ref{fig5}, where the working frequency is at deep sub-wavelength scale. We have demonstrated that only evanescent modes can exist at the ST type edge. The two \gls{pc}s in Fig.\ref{fig7} are designed to have only LT type edges. Based on our description, neither traveling nor evanescent waves are supported along these edges. As a result, when the topological wave arrives at the very end of our \gls{pc}s, the resonators undergo strong local resonances. For the symmetric valley modes traveling along SDW, the in-phase resonance radiates outward a monople field, as shown in Fig.\ref{fig7}(a). While for the antisymmetric modes in Fig.\ref{fig7}(b), the out-of-phase local resonances act like a dipole source.

\section{Conclusions \label{sec:V}}

In conclusion, we have proposed an elastic valley \gls{pc} by emulating \gls{qvhe}, achieved simply by tuning the mass perturbation of the resonators. We have shown that a biharmonic equation  near the valley points can be mapped into the massive Dirac Hamiltonian by using the $\mathbf{k}\cdot\mathbf{p}$ method. Moreover, the mass perturbation does indeed contribute to the Dirac mass term, which is responsible for the topological band gap opening. Two different types of domain walls can be built when constructing the valley \gls{pc}. Each of them supports a topological symmetric and antisymmetric valley modes, respectively. In contrast with the S mode that refracts into a single wave beam, the A mode refracts into two out-of-phase beams through an armchair outlet edge because of its unique horizontal valley splitting. By parameterize the average mass $\gamma$ and the perturbation $\Delta\gamma$, the decay depth of these localized modes into the bulk is found to be determined solely by the ratio $|\Delta\gamma|/\gamma$. Moreover, large mass perturbation would result in strong resonance between scatterers near the domain wall, and may transform the refracted plane wave into a monopole and dipole-like radiated fields, for S and A valley modes respectively. Furthermore, increasing $\gamma$ to a large value enables going down to deep sub-wavelength scale, where the topological refracted wave becomes evanescent and can only propagate along a specific type of edge. Finally, we have presented two examples of topological protected functional devices designs, including an elastic diode and a near perfect monople/dipole emitter.

\section*{ACKNOWLEDGEMENTS}
Linyun Yang acknowledges the research Grant No. 201906120081
provided by the China Scholarship Council (CSC). Ying Wu acknowledges the support from China Postdoctoral Science Foundation (NO. 2020M672615).








\bibliography{References}

\end{document}